# SU($N$) toric code and non-Abelian anyons

Manu Mathur[*] and Atul Rathor[†]

*S. N. Bose National Centre for Basic Sciences, Salt Lake, Kolkata 700 106, India*

We construct an SU($N$) toric code model describing the dynamics of SU($N$) electric and magnetic fluxes on a two-dimensional torus. We show that the model has $N^2$ topologically distinct ground states $|\psi_0\rangle_{(p,q)}$, which are the loop states characterized by $Z_N \otimes Z_N$ center charges (p, q = 0, 1, 2, . . . , $N - 1$). We explicitly construct them in terms of coherent superpositions of all possible spin network states on a torus with Wigner coefficients as their amplitudes. All excited quasiparticle states with SU($N$) electric charges and magnetic fluxes are constructed. We show that the braiding statistics of these SU($N$) electric or magnetic quasiparticles or non-Abelian anyons are encoded in the Wigner rotation matrices.

## I. INTRODUCTION

In the past few years topological quantum computing with Abelian and non-Abelian anyons has become one of the most promising approaches for storing and processing quantum information reliably in systems on two-dimensional surfaces [1–3]. In these computational approaches, the quantum information is encoded in the topological braiding of anyons, therefore making it resilient against any local perturbations or decoherence effects. All these models with anyons originated from the simple $Z_2$ toric code model proposed by Kitaev [1]. This (2 + 1)-dimensional model is an exactly solvable model with four degenerate ground states which are characterized by $Z_2 \otimes Z_2$ topological order. The anyonic excitations in this model and the topological nature of their braiding have led to the idea of topological or fault-tolerant quantum computers having an intrinsic resistance to small perturbations. In the past, the appealing property of topological ordering and quantum error correction of the $Z_2$ toric code model has led to other interesting models with discrete Abelian and non-Abelian groups [4]. Kitaev himself proposed a model of quantum doubles [1]. Levin and Wen introduced their string net models with string net condensation as the basic mechanism underlying the topological ordering [5].

In this work we generalize Kitaev's $Z_2$ toric code model to SU($N$) group leading to non-Abelian anyons. In the recent past models with non-Abelian anyons have been subject of intense research for their quantum computing applications [2,3]. The SU($N$) toric code model, discussed in this work, provides a natural setting for such non-Abelian exotic quasiparticles as electric and magnetic excitations over the ground states. This SU($N$) model has $N^2$ degenerate ground states which are loop or spin network states and are characterized by $Z_N \otimes Z_N$ topological charges. They are explicitly constructed in terms of spin networks and the Wigner coefficients as their amplitudes.

In fact, besides their topological stability, the ground states of the SU($N$) toric code model are also geometrically rigid because of the numerous interlinked triangular electric flux constraints satisfied by the underlying spin network states[1] over the entire torus. This geometrical stability is purely a group-theoretic property of the Hilbert space of the SU($N$) toric code model. The spin networks, constructed in Sec. III A, are similar to the string nets of Levin and Wen which were discussed in the context of topological phases and non-Abelian anyons [5]. Both string nets and spin networks represent the physical Hilbert spaces after solving the Gauss law constraints. We also construct all excited states using SU($N$) link holonomies and the non-Abelian vortex creation-annihilation operator. We show that the SU($N$) canonical commutation relations between electric fields and the conjugate potentials lead to the non-Abelian anyonic nature of these excitations or quasiparticles. We further show that their mutual non-Abelian statistics are encoded in Wigner $D$ matrices.

The organization of the paper is as follows. In Sec. II we briefly discuss the kinematic aspects of SU($N$) lattice gauge theory in the Hamiltonian formulation. The purpose of this section is to make the presentation self-contained. More details can be found in the excellent original review article in [6]. In Sec. III we construct the SU($N$) toric code model and the associated operators and their algebras are discussed. In Sec. III A we construct the ground state in the topologically trivial sector. In Sec. III A 1 we extend this construction to include $Z_N \otimes Z_N$ topological charges, where $Z_N$ is the center of the group SU($N$). Section III B discusses the excited states of the SU($N$) toric code Hamiltonian. In Sec. III B 1 we discuss excitations with non-Abelian electric charges. In Sec. III B 2 we discuss the non-Abelian magnetic excitations and the corresponding vortex operators [7]. In Sec. IV we discuss the non-Abelian anyonic nature of the above excitations.

---

[*]manu@bose.res.in; manumathur14@gmail.com
[†]atulrathor@bose.res.in; atulrathor999@gmail.com

---

[1]In the SU(2) case, there are $2L^2$ triangular constraints on a torus with $L^2$ lattice sites. The $L = 2$ case with eight triangular constraints on the torus is worked out completely (see Sec. III A and Fig. 3).

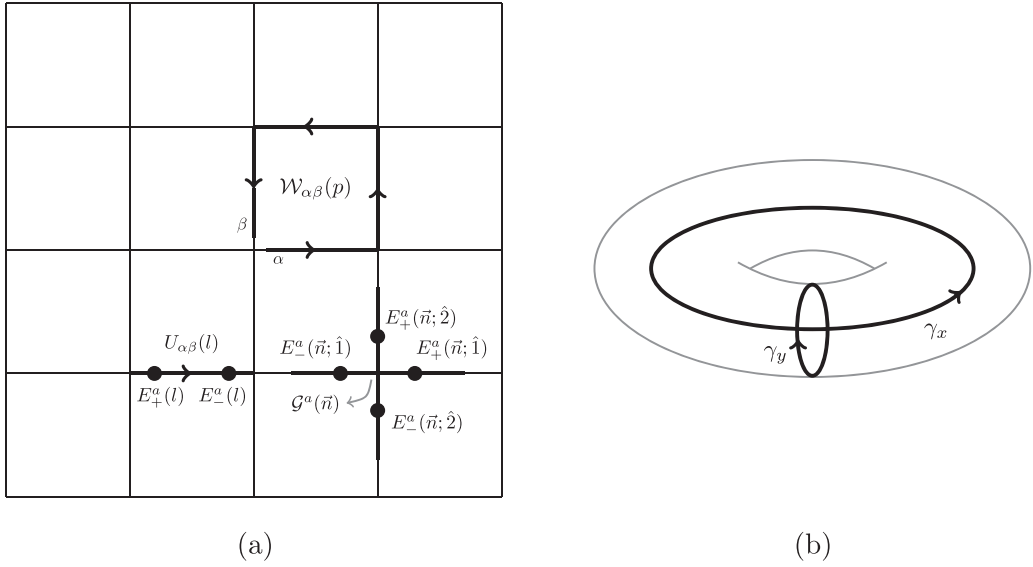

FIG. 1. (a) The SU(N) electric fields $E_\pm^a(l)$ and the conjugate link operators $U_{\alpha\beta}(l)$. The operators $\mathcal{G}^a(\vec{n})$ and $\mathcal{W}(p)$ appearing in the Hamiltonian (9) are also shown. (b) Two noncontractible Wilson loops $\gamma_x$ and $\gamma_y$ on the torus $\mathcal{T}_2$.

Throughout this paper, we use the following notation. The lattice sites on the two-dimensional torus, denoted by $\mathcal{T}_2$, are labeled as $\vec{n} = (x, y), x, y = 1, 2, \ldots, L$. The lattice links are denoted by $l \equiv (\vec{n}; \hat{i})$, with $i = 1, 2$. We will often denote the sites by $v$ or $s$, links by $l$, and plaquettes by $p$ for simplicity if other details are not required. On the torus $\mathcal{T}_2$ there are $L^2$ sites, $2L^2$ links, and $L^2$ plaquettes. The periodicities in two directions imply $(x + L, y) = (x, y + L) = (x, y)$. We will often use the SU(2) group for explicit calculations. This is just to avoid unnecessary group-theoretic technicalities involved with SU(N), $N \geqslant 3$.

## II. SU(N) HAMILTONIAN FORMULATION

The kinematic variables involved in Kogut and Susskind's Hamiltonian formulation [6] of lattice gauge theories are SU(N) link operators $U(\vec{n}, \hat{i}) \equiv U(l)$ and the corresponding conjugate link electric fields $E_+^a(\vec{n}, \hat{i}) \equiv E_+^a(l)$ and $E_-^a(\vec{n} + \hat{i}; \hat{i}) \equiv E_-^a(l)$. These electric fields, shown in Fig. 1(a), rotate the link holonomies $U(l)$ from the left and right ends, respectively, leading to the canonical commutation relations

$$[E_+^a(l), U_{\alpha\beta}(l)] = [T^a U(l)]_{\alpha\beta},$$
$$[E_-^a(l), U_{\alpha\beta}(l)] = -[U(l) T^a]_{\alpha\beta}. \quad (1)$$

In (1), $T^a$ are the generators in the fundamental representation of SU(N) satisfying $\text{Tr}(T^a T^b) = \frac{1}{2}\delta_{ab}$. The above commutation relations, along with the Jacobi identities $[E_\pm^a(l), [E_\pm^b(l), U_{\alpha\beta}(l)]] + (2 \text{ cyclic permutations}) = 0$, yield [6]

$$[E_+^a(l), E_+^b(l)] = i f^{abc} E_+^c(l),$$
$$[E_-^a(l), E_-^b(l)] = -i f^{abc} E_-^c(l). \quad (2)$$

In (2), $f^{abc}$ are the SU(N) structure constants. The left and right electric fields in (1) are related by a parallel transport [6]:

$$E_-(l) = -U^\dagger(l) E_+(l) U(l). \quad (3)$$

In (3) we have defined $E_\pm(l) \equiv \sum_a E_\pm^a(l) T^a$. It is easy to check that on any link $l = (\vec{n}; \hat{i})$, the magnitudes of the left and right electric fields are equal and they commute with each other:

$$\text{Tr}[\vec{E}_+(l) \cdot \vec{E}_+(l)] = \text{Tr}[\vec{E}_-(l) \cdot \vec{E}_-(l)] \equiv \text{Tr}\vec{E}^2(l), \quad (4a)$$
$$[\vec{E}^2(l), E_+^a(l)] = 0, \quad [\vec{E}^2(l), E_-^a(l)] = 0, \quad (4b)$$
$$[E_+^a(l), E_-^b(l)] = 0. \quad (4c)$$

The important identities (4a), (4b), and (4c) imply that the Hilbert space in the simplest SU(2) case is spanned by the eigenvectors $|j(l), m_+(l), m_-(l)\rangle \equiv |j, m_+, m_-\rangle_l = |j, m_+\rangle_l \otimes |j, m_-\rangle_l$ on every link $l$ satisfying

$$[\vec{E}_\pm(l)]^2 |j, m_+, m_-\rangle_l = j(l)[j(l) + 1]|j, m_+, m_-\rangle_l,$$
$$E_\pm^{a=3}(l)|j, m_+, m_-\rangle_l = m_\pm(l)|j, m_+, m_-\rangle_l. \quad (5)$$

The SU(N) transformations[2] on the flux operators $U(\vec{n}; \hat{i})$ and electric fields $E_\pm(n; \hat{i})$ are [6]

$$U(\vec{n}; \hat{i}) \to \Lambda(\vec{n}) U(\vec{n}; \hat{i}) \Lambda^\dagger(\vec{n} + \hat{i}),$$
$$E_\pm(\vec{n}; \hat{i}) \to \Lambda(\vec{n}) E_\pm(\vec{n}; \hat{i}) \Lambda^\dagger(\vec{n}), \quad i = 1, 2. \quad (6)$$

In (6), $\Lambda(\vec{n})$ are the SU(N) matrices describing $N^2 - 1$, SU(N) rotational degrees of freedom at every lattice site $\vec{n}$. The canonical commutation relations (1) imply that the generators of the above SU(N) rotations at site $\vec{n}$ are

$$\mathcal{G}^a(\vec{n}) \equiv \sum_{i=1}^{2} [E_+^a(\vec{n}; \hat{i}) + E_-^a(\vec{n}; \hat{i})]. \quad (7)$$

---

[2]The transformations (6) are the SU(N) symmetries of the toric code Hamiltonian (9). They do not correspond to the redundancies in SU(N) gauge theory which are removed by the Gauss law constraints $\mathcal{G}^a(n) = 0$ at every lattice site.

The Gauss law operators $\mathcal{G}^a(n)$ are illustrated in Fig. 1(a). We also define SU(*N*) plaquette operators $\mathcal{W}(p) \equiv \mathcal{W}(\vec{n})$ which contain SU(*N*) magnetic fields[3] [see Fig. 1(a)]

$$\mathcal{W}_{\alpha\beta}(p) = [U(\vec{n};\hat{1})U(\vec{n}+\hat{1};\hat{2})U^\dagger(\vec{n}+\hat{2};\hat{1})U^\dagger(\vec{n};\hat{2})]_{\alpha\beta}. \quad (8)$$

## III. SU(*N*) TORIC CODE MODEL

We now generalize Kitaev's toric code Hamiltonian and write the SU(*N*) toric code Hamiltonian as

$$H = A \sum_n A_n + B \sum_p B_p. \quad (9)$$

In the above Hamiltonian *n* and *p* denote the sites and plaquettes, respectively, on the torus $\mathcal{T}_2$; *A* and *B* are positive constants; and

$$A_n \equiv \sum_{a=1}^{N^2-1} \mathcal{G}^a(n)\mathcal{G}^a(n), \quad (10)$$

$$B_p \equiv 1 - \frac{1}{2N}\text{Tr}[\mathcal{W}(p) + \mathcal{W}^\dagger(p)]. \quad (11)$$

Under SU(*N*) transformations (6) both electric and magnetic field terms in (7) and (8), respectively, transform covariantly

$$\mathcal{G}(n) \to \Lambda(n)\mathcal{G}(n)\Lambda^\dagger(n),$$
$$\mathcal{W}(p) \to \Lambda(n)\mathcal{W}(p)\Lambda^\dagger(n). \quad (12)$$

In (12), $\mathcal{G} \equiv \sum_a \mathcal{G}^a T^a$. Therefore, the toric code Hamiltonian in (9) is invariant under the SU(*N*) rotations (6). Like Kitaev's model, the Hamiltonian (9) is a sum of $L^2$ electric terms and $L^2$ magnetic terms and they all commute with each other[4]

$$[A_n, A_{n'}] = 0, \quad [A_n, B_{p'}] = 0,$$
$$[B_p, B_{p'}] = 0 \,\forall\, n, n', p, p'. \quad (13)$$

On a torus we have additional topological invariants defined over noncontractible loops. These SU(*N*) Wilson loop operators in the fundamental representation are

$$W_{\gamma_x} = \text{Tr}\prod_{l\in\gamma_x} U(l), \quad W_{\gamma_y} = \text{Tr}\prod_{l\in\gamma_y} U(l). \quad (14)$$

In (14), $\gamma_x$ and $\gamma_y$ are the two independent oriented paths shown in Fig. 1(b) and the products over links are path ordered. The paths $\gamma_x$ and $\gamma_y$ start from an arbitrary base point $(x_0, y_0)$ and return to it after looping the torus in the horizontal and vertical directions, respectively. It is easy to check that they satisfy

$$[W_{\gamma_x}, W_{\gamma_y}] = 0, \quad [W_{\gamma_x}, A_n] = 0, \quad [W_{\gamma_x}, B_p] = 0,$$
$$[W_{\gamma_y}, A_n] = 0, \quad [W_{\gamma_y}, B_p] = 0. \quad (15)$$

Note that, unlike the $\mathbb{Z}_2$ toric code model, we do not have simple ground-state projector operators for the present SU(*N*) generalization.[5] In the next section we will exploit (13) and (15) to characterize the topological sectors of the SU(*N*) toric code model and construct the $N^2$ ground states with topological charges.

### A. Ground states

We first study the ground-state structure of the SU(*N*) toric code Hamiltonian (9). A ground state $|\psi_0\rangle$ satisfies

$$A_n|\psi_0\rangle = 0, \quad B_p|\psi_0\rangle = 0 \,\forall\, n, p \in \mathcal{T}_2. \quad (16)$$

The first condition in (16) enforces SU(*N*) rotational invariance on the ground state $|\psi_0\rangle$. These constraints are the same as $N^2-1$, SU(*N*) Gauss law constraints in SU(*N*) lattice gauge theory [6–8]. Therefore, all possible ground states of the SU(*N*) toric code Hamiltonian are all mutually independent loop states of the SU(*N*) lattice gauge theory with $B_p = 0$. These loop states are best analyzed and characterized in the dual electric flux basis [7,8]. We note that in two space dimensions there are four SU(*N*) electric fluxes meeting at a vertex. Their quantum numbers can be characterized by the corresponding four SU(*N*) Young tableaux. The Gauss law constraint $A_v = 0$ states that the ground-state manifold consists of all possible independent SU(*N*) invariant states with total SU(*N*) electric flux equal to zero. In the simple $N = 2$ case [8], a basis in the loop Hilbert space at a vertex *s* is given by $|j_1, j_2, j_{12} = j_{34}, j_3, j_4\rangle_s \equiv |\vec{J}\rangle_s$. This is shown in Fig. 2(c). They satisfy

$$\mathcal{G}^a(s)|\vec{J}\rangle_s = 0 \,\forall\, s \in \mathcal{T}_2. \quad (17)$$

Therefore, if we label the $L^2$ vertices or sites on the torus by $v_1, v_2, \ldots, v_{L^2}$, then the ground state of the SU(2) toric code is of the form

$$|\psi_0\rangle = \sum_{\{\vec{J}_{v_1},\vec{J}_{v_2},\ldots\}} \underbrace{W(\vec{J}_{v_1},\vec{J}_{v_2},\ldots,)}_{\text{amplitude on torus}} \underbrace{\left\{\prod_{i=1}^{L^2} \otimes|\vec{J}\rangle_{v_i}\right\}}_{\text{loop states on torus}}. \quad (18)$$

In (18) the summation over $\{\vec{J}_{v_i}\}$ is constrained as each electric flux $j(l)$ is shared by two vertices at the two ends of the link *l*. The first ground-state condition $A_v|\psi_0\rangle = 0$ is ensured by the spin network identities (17) for arbitrary amplitudes $W(\vec{J}_{v_1},\vec{J}_{v_2},\ldots)$. These amplitudes are now fixed by demanding $B_p|\psi_0\rangle = 0$. However, we find this approach difficult as the action of $B_p$ on a spin network state is not simple [8]. We therefore reverse the process and start with completely ordered magnetic eigenstates with $B_p = 0 \,\forall\, p$ and then ensure $A_v|\psi_0\rangle = 0$ by demanding invariance under the symmetry transformations (6) at every vertex. We first notice that the eigenstates of $B_p$ in (11) are necessarily eigenstates of the individual link holonomies $U(l)_{\alpha\beta}$. As all these flux operators commute with each other $[U_{\alpha\beta}(l), U_{\gamma\delta}(l')] = 0 \, \forall\, l, l'$ and $\alpha, \beta, \gamma, \delta$, we can diagonalize all of them simultaneously. Further, the eigenvalues of any SU(2) matrix operator $U_{\alpha\beta}(l)$ are SU(2) matrices; therefore, we define the SU(2) group manifold $S^3$ on every link $l$:

$$Z(l) = \begin{bmatrix} z_1(l) & z_2(l) \\ -z_2^*(l) & z_1^*(l) \end{bmatrix}. \tag{19}$$

---

[3]The SU(*N*) magnetic fields on plaquette *p* are defined as $B^a(p) \equiv \frac{1}{2}\text{Tr}T^a[2 - \mathcal{W}(p) - \mathcal{W}^\dagger(p)]$.

[4]The first two relations in (13) are the symmetry statements of the toric code Hamiltonian (9). The first equation follows from the fact that $[\mathcal{G}^a(n), \mathcal{G}^a(n')] = 0 \,\forall\, n, n'$ because of the identity (4c). The second relation is because $B_p$ is invariant under SU(*N*) rotations (6). The third relation is trivially true because all flux operators $U_{\alpha\beta}(l)$ commute among themselves.

[5]The $\mathbb{Z}_2$ toric code model has simple ground-state projectors (see the work of Schulz *et al.* in [4]) $\mathcal{P} \equiv \prod_n[1 + \sigma_1(l_1)\sigma_1(l_2)\sigma_1(l_3)\sigma_1(l_4)] \prod_p[1 + \sigma_3(l_1)\sigma_3(l_2)\sigma_3(l_3)\sigma_3(l_4)]$, where $l_1, l_2, l_3, l_4$ are the edges sharing the site *n* in the first term and denote the boundaries of the plaquette *p* in the second term. Such projectors do not exist in the SU(*N*) toric code model.

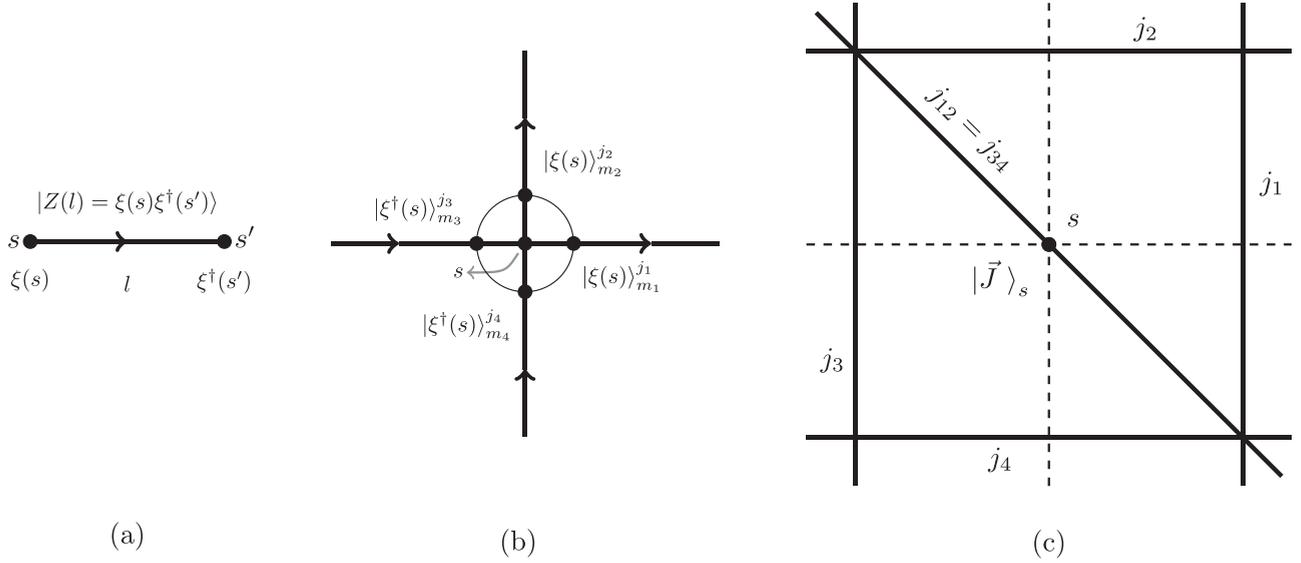

FIG. 2. Ground states of the SU(2) toric code, from (a) link states $|Z(l)\rangle$ to (b) site states $|\xi(s)\rangle_m^j$ to (c) loop states $|\vec{J}\rangle_s \equiv |j_1, j_2, j_{12} = j_{34}, j_3, j_4\rangle_s$. The geometrical triangular constraints manifest on the dual lattice (solid lines) around the site $s$. Every lattice site has two triangular constraints: $\{j_1, j_2, j_{12}\}$ and $\{j_3, j_4, j_{34} = j_{12}\}$ must form triangles.

In (19), $(z_1, z_2) \in S^3$ and satisfy $|z_1(l)|^2 + |z_2(l)|^2 = 1$ and $Z(-l) \equiv Z^\dagger(l) \, \forall\, l$. We define the magnetic eigenstates to be eigenstates of each link operator:

$$U_{\alpha\beta}(l)|Z(l)\rangle = Z_{\alpha\beta}(l)|Z(l)\rangle. \tag{20}$$

The eigenvectors are

$$|Z(l)\rangle \equiv |z_1(l), z_2(l)\rangle$$
$$= \frac{1}{4\pi} \sum_{j=0}^{\infty} \sqrt{(2j+1)} \sum_{m_\pm} D^j_{m_+, m_-}[Z(l)]$$
$$\times |jm_+\rangle_l \otimes |jm_-\rangle_l. \tag{21}$$

To obtain the ground state with $B_p = 0$ or equivalently $\mathcal{W}_{\alpha\beta}(p) = \delta_{\alpha\beta}$, we choose pure gauge conditions on every link to write

$$Z(l) \equiv Z(\vec{n}, \hat{l}) = \xi(s)\xi^\dagger(s'), \tag{22}$$

as shown in Fig. 2(a). At each site $s \in \mathcal{T}_2$, $(\xi(s) \equiv \xi_1(s), \xi_2(s)) \in S^3$, with $|\xi_1(s)|^2 + |\xi_2(s)|^2 = 1$, and can be written as the SU(2) matrix (19). We can thus write the maximally ordered $(\mathcal{W}_p = 1 \, \forall\, p)$ magnetic eigenstates in (21) as the direct product of site states defined at the two end sites of the link $l$,

$$|Z(l)\rangle = \sum_{j=0}^{\infty} \sum_{m=-j}^{j} |\xi(s)\rangle_m^j \otimes |\xi^\dagger(s')\rangle_m^j. \tag{23}$$

The site states at the left and right ends ($s$ and $s'$) of the link $l$ are defined as

$$|\xi(s)\rangle_m^j = \sqrt{d_j} \sum_{m_+} D^j_{m_+,m}[\xi(s)]|j, m_+\rangle_l,$$
$$|\xi^\dagger(s')\rangle_m^j = \sqrt{d_j} \sum_{m_-} D^j_{m,m_-}[\xi^\dagger(s')]|j, m_-\rangle_l. \tag{24}$$

Above $d_j \equiv \sqrt{(2j+1)}/4\pi$. Therefore, all states satisfying $B_p|\psi\rangle = |\psi\rangle$ can be written as product of the site states at every site $s$,

$$|\psi\rangle = \sum_{\text{all } j} \sum_{\text{all } m} \prod_{s \in \mathcal{T}_2} |\xi(s)\rangle_{m_1}^{j_1} \otimes |\xi(s)\rangle_{m_2}^{j_2}$$
$$\otimes |\xi^\dagger(s)\rangle_{m_3}^{j_3} \otimes |\xi^\dagger(s)\rangle_{m_4}^{j_4}. \tag{25}$$

We can now perform the integrations over $\xi(s)$ at every site $s$ to get the loop states satisfying both conditions in (16). Using the multiplicative properties of the Wigner rotation matrices [9], these group manifold integrations at every lattice site can be performed exactly to get

$$|\psi_0\rangle = \sum_{\text{all } j} \sum_{\text{all } m} \prod_{s \in \mathcal{T}_2} \left\{ \int_{S^3} d^2\mu[\xi(s)] |\xi(s)\rangle_{m_1}^{j_1} \right.$$
$$\left. \otimes |\xi(s)\rangle_{m_2}^{j_2} \otimes |\xi^\dagger(s)\rangle_{m_3}^{j_3} \otimes |\xi^\dagger(s)\rangle_{m_4}^{j_4} \right\}$$
$$= \sum_{\{\vec{J}\}} W\{\vec{J}\} \prod_s \otimes \underbrace{|j_1, j_2, j_{12} = j_{34}, j_3, j_4\rangle_s}_{\text{loop state at site } s}$$
$$= \sum_{\{\vec{J}\}} W\{\vec{J}\} \prod_s \otimes |\vec{J}\rangle_s. \tag{26}$$

In (26) the loop states at every site $s$ are

$$|\vec{J}\rangle_s = \sum_{\text{all } m's} \eta^{j_{12}}_{m_{12}} C^{j_{12} m_{12}}_{j_1 m_1, j_2 m_2} C^{j_{12}, -m_{12}}_{j_3 m_3, j_4 m_4} \prod_{i=1}^{4} \otimes |j_i m_i\rangle. \tag{27}$$

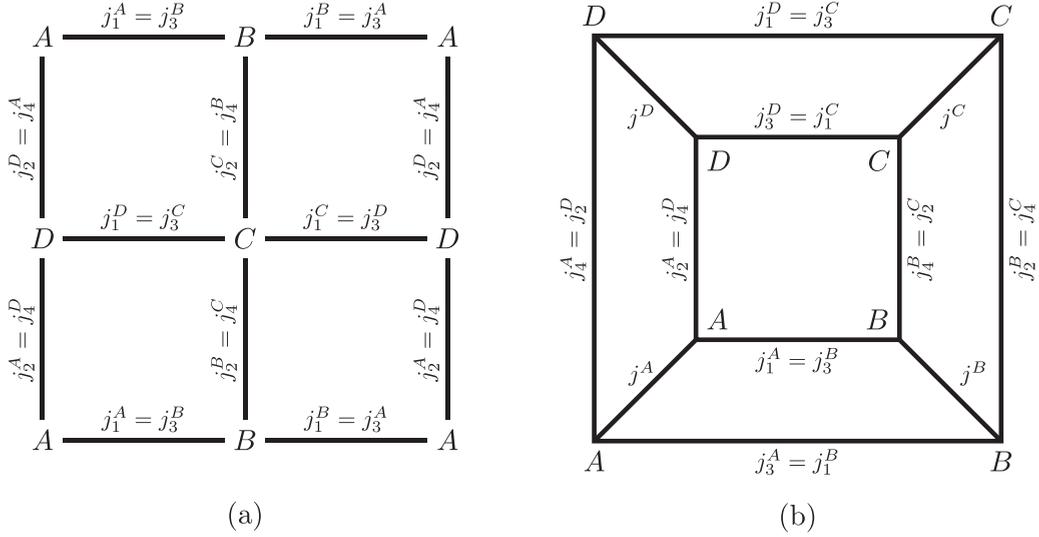

FIG. 3. SU(2) toric code. (a) Electric fluxes on a four-plaquette toric code and (b) 12-*j* Wigner coefficients involving eight triangular constraints as the amplitude for the SU(2) toric code ground state.

In (27), $\eta_m^j \equiv (-1)^{j-m}(2j+1)^{-1/2}$ and the loop states $|\vec{J}\rangle_s \equiv |j_1, j_2, j_{12} = j_{34}, j_3, j_4\rangle_s$ contain two group-theoretic triangular constraints [8] around every lattice site. They are illustrated on the dual lattice plaquette in Fig. 2(c). The amplitudes $W\{\vec{J}\}$ can be computed after integrating over all $L^2$ lattice sites and then summing over the remaining magnetic quantum numbers. We note that all the magnetic quantum numbers are completely contained in the Clebsch-Gordan coefficients appearing as the coefficients of the loop states

$$\int_{S^3} d^2\mu[\xi(s)]|\xi(s)\rangle_{m_1}^{j_1} \otimes |\xi(s)\rangle_{m_2}^{j_2} \otimes |\xi^\dagger(s)\rangle_{m_3}^{j_3} \otimes |\xi^\dagger(s)\rangle_{m_4}^{j_4}$$
$$= K_{\{j_1,j_2,j_3,j_4\}}(-1)^{j_3-m_3+j_4-m_4}\sum_{j_{12},m_{12}} \eta_{m_{12}}^{j_{12}} C_{j_1,m_1 j_2,m_2}^{j_{12},m_{12}}$$
$$\times C_{j_3,m_3 j_4,m_4}^{j_{12},-m_{12}}|j_1, j_2, j_{12}, j_3, j_4\rangle_s,$$

where $K_{\{j_1,j_2,j_3,j_4\}} = \prod_{\{\text{all } j\}}(2j+1)^{1/4}$. The sums over the remaining magnetic quantum numbers ($m_1, m_2, m_3, m_4$) attached to every site can be done after integration over all $L^2$ lattice sites on $\mathcal{T}_2$. These amplitudes ensure that the triangular flux constraints are satisfied at every lattice site and they are also glued together on the entire torus $\mathcal{T}_2$. It is illustrative to consider a simple example and understand this construction. We consider a torus with four lattice sites $A$, $B$, $C$, and $D$ and four associated plaquettes as shown in Fig. 3(a). The ground state in (26) can now be written as

$$|\psi_0\rangle = \sum_{\{\vec{J}_A, \vec{J}_B, \vec{J}_C, \vec{J}_D\}} \underbrace{W(\vec{J}_A, \vec{J}_B, \vec{J}_C, \vec{J}_D)}_{\text{amplitude}}$$
$$\times \underbrace{|\vec{J}_A\rangle \otimes |\vec{J}_B\rangle \otimes |\vec{J}_C\rangle \otimes |\vec{J}_D\rangle}_{\text{loop states on torus}}. \quad (28)$$

The amplitudes $W(\vec{J}_A, \vec{J}_B, \vec{J}_C, \vec{J}_D)$ in (28) are the 12-*j* Wigner coefficients of the second kind:

$$W(\vec{J}_A, \vec{J}_B, \vec{J}_C, \vec{J}_D) = \Pi \begin{bmatrix} j_4^D = j_2^A & j_1^A = j_3^B & j_4^B = j_2^C & j_1^C = j_3^D \\ & j^A & & j^B & & j^C & & j^D \\ j_2^D = j_4^A & j_3^A = j_1^B & j_2^B = j_4^C & j_3^C = j_1^D \end{bmatrix}. \quad (29)$$

We have defined $\Pi = \prod_{(\text{all } j)}\sqrt{(2j+1)}$. The 12-*j* Wigner coefficients are shown in Fig. 3(b). We thus have $A_v|\psi_0\rangle = 0$ because of the spin network structure in (27) and $B_p|\psi_0\rangle = 0$ because of the 12-*j* coefficients in (29). The pure gauge conditions (22) make the ground state $|\psi_0\rangle$ satisfy

$$W_{\gamma_x}|\psi_0\rangle = |\psi_0\rangle, \quad W_{\gamma_y}|\psi_0\rangle = |\psi_0\rangle. \quad (30)$$

Hence $|\psi_0\rangle$ is in the trivial topological sector. We now construct states with topological charges.

### 1. $Z_N \otimes Z_N$ charges

To construct topologically nontrivial ground states carrying $Z_N \otimes Z_N$ charges, we generalize the pure gauge conditions (22) as follows. We attach $Z_N \otimes Z_N$ phase factors $\eta_x$ and $\eta_y$ to all the links along the two strips $\mathcal{S}_y$ and $\mathcal{S}_x$ encircling the torus as shown in Figs. 4(a) and 4(b):

$$Z(l_x) = \xi(s)\eta(l_x)\xi^\dagger(s'), \quad \eta(l_x) \equiv \eta_x = e^{2\pi i \mathsf{p}/N}, \quad l_x \in \mathcal{S}_y,$$
$$Z(l_y) = \xi(s)\eta(l_y)\xi^\dagger(s'), \quad \eta(l_y) \equiv \eta_y = e^{2\pi i \mathsf{q}/N}, \quad l_y \in \mathcal{S}_x. \quad (31)$$

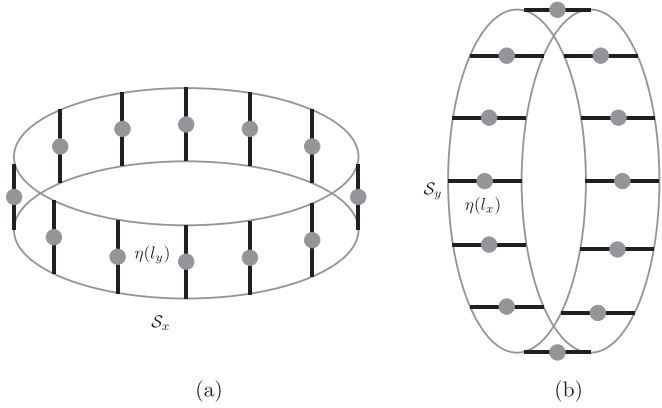

FIG. 4. Construction of topologically nontrivial ground states, inserting $Z_N$ twist factors, shown as •, in the ground states: (a) $Z_N$ phase $\eta_y = e^{2\pi i q/N}$ along $\mathcal{S}_x$ and (b) $Z_N$ phase $\eta_x = e^{2\pi i p/N}$ along $\mathcal{S}_y$.

In (31), $\mathsf{p}, \mathsf{q} = 0, 1, 2, \ldots, N-1$. Repeating the procedure of the preceding section after replacing (22) with (31), we get

$$|\psi_0\rangle_{(\mathsf{p},\mathsf{q})} = \sum_{\{\vec{J}\}} (\eta_x)^{\{\mathcal{J}_x\}} (\eta_y)^{\{\mathcal{J}_y\}} W\{\vec{J}\}$$
$$\times \prod_s \otimes \underbrace{|j_1, j_2, j_{12} = j_{34}, j_3, j_4\rangle_s}_{\text{loop state at site } s}. \quad (32)$$

The additional phase factors in (32) leading to topologically nontrivial sectors are

$$\mathcal{J}_x = \sum_{l_x \in \mathcal{S}_y} j(l_x), \quad \mathcal{J}_y = \sum_{l_y \in \mathcal{S}_x} j(l_y). \quad (33)$$

The ground states $|\psi_0\rangle_{(\mathsf{p},\mathsf{q})}$ satisfy

$$W_{\gamma_x}|\psi_0\rangle_{(\mathsf{p},\mathsf{q})} = \eta_x|\psi_0\rangle_{(\mathsf{p},\mathsf{q})},$$
$$W_{\gamma_y}|\psi_0\rangle_{(\mathsf{p},\mathsf{q})} = \eta_y|\psi_0\rangle_{(\mathsf{p},\mathsf{q})}. \quad (34)$$

The ground state in the trivial sector in (26) is $|\psi_0\rangle = |\psi\rangle_{(\mathsf{p}=0,\mathsf{q}=0)}$. Having constructed all $N^2$ ground states, we now discuss the quasiparticle excited states and then their non-Abelian anyonic nature.

### B. Non-Abelian electric and magnetic excitations

#### 1. Electric fluxes

We now define the Wilson charge operator

$$\Gamma^{(N)}_{\alpha\beta}(n', n) = \left(\prod_{l \in \mathcal{L}}^p U(l)\right)_{\alpha\beta}. \quad (35)$$

In (35), $\prod^p$ denotes the lattice path ordered product of link holonomies in the SU(N) fundamental representation along the oriented path $\mathcal{L}$ as shown in Fig. 5. The string operator is invariant under rotations (6) all along $\mathcal{L}$ except at the end point $n$ ($n'$), where it creates SU(N) quasiparticle in the fundamental $N$ (antifundamental $\bar{N}$) representation:

$$[A_v, \Gamma^{(N)}_{\alpha\beta}(n', n)] = \begin{cases} \frac{(N^2-1)}{N^2} \Gamma^{(j)}_{\alpha\beta}(n', n) & \text{if } v = n \text{ or } n' \\ 0 & \text{otherwise.} \end{cases} \quad (36)$$

We also note that the length and shape of the string $\mathcal{L}$ between the end points are invisible and therefore unphysical. As the $U_{\alpha\beta}(l)$ commute among themselves,

$$[B_p, \Gamma^{(N)}_{\alpha\beta}(n', n)] = 0 \; \forall \; p; n', n. \quad (37)$$

Thus the quasiparticle states

$$\left|\psi^{(N)}_{\alpha\beta}(n', n)\right\rangle_{(\mathsf{p},\mathsf{q})} \equiv \Gamma^{(N)}_{\alpha\beta}(n', n)|\psi_0\rangle_{(\mathsf{p},\mathsf{q})} \quad (38)$$

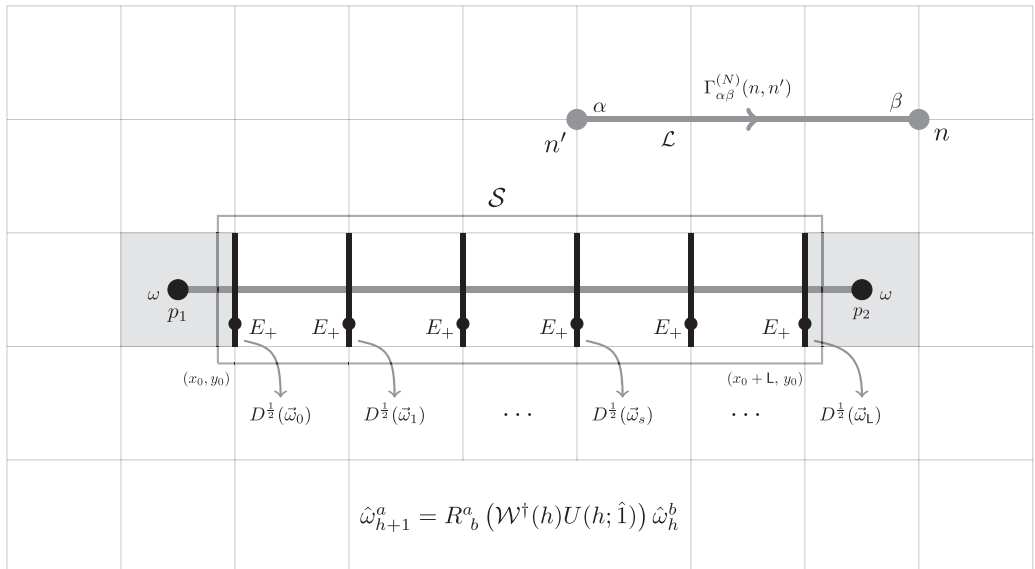

FIG. 5. Creating electric charges at sites and magnetic vortices on plaquettes. The operator $\Gamma^{(N)}_{\alpha\beta}(n', n)$ creates electric charges in the fundamental representations at the two end points $n$ and $n'$. The operator $\Sigma_{\vec{\omega}}$ acts on all the vertical link fluxes $U(s; \hat{2}) \in \mathcal{S}$ (dark vertical links) and rotates all of them by the same angle $\omega(x_0, y_0) = \omega_{s=0} \equiv \omega$. However, the rotation axes $\hat{\omega}_{s=1}, \hat{\omega}_{s=2}, \ldots, \hat{\omega}_{s=L}$ are different and related to $\hat{\omega}_0$ by parallel transports (42). Note that the shapes and lengths of $\mathcal{L}$ and $\mathcal{S}$ are unphysical.

are the eigenstates of the SU(*N*) toric code Hamiltonian (9),

$$H|\psi^{(N)}_{\alpha\beta}(n',n)\rangle_{(\mathsf{p},\mathsf{q})} = 2A\frac{N^2-1}{N^2}|\psi^{(N)}_{\alpha\beta}(n',n)\rangle_{(\mathsf{p},\mathsf{q})}. \quad (39)$$

We have obtained the above eigenvalue equation using (36). Therefore, $|\psi^{(N)}_{\alpha\beta}(n',n)\rangle_{(\mathsf{p},\mathsf{q})}$ are the eigenstates of $H$ representing quasiparticles in the fundamental (antifundamental) representations at lattice sites $n$ ($n'$). We can similarly construct electric states in the higher representations of SU(*N*). These state will be characterized by the corresponding SU(*N*) Young tableau.

### 2. Magnetic fluxes

We now construct unitary vortex operators which create and destroy non-Abelian magnetic fluxes on two plaquettes [7]. The SU(2) magnetic fluxes on the plaquettes in the axis angle representation $\vec{\omega} \equiv (\hat{\omega}, \omega)$ can be written as

$$\mathcal{W} = \cos\left(\frac{\omega}{2}\right)\sigma_0 + i\hat{\omega}\cdot\vec{\sigma}\sin\left(\frac{\omega}{2}\right). \quad (40)$$

To construct a vortex or disorder operator creating the above magnetic flux, we consider a finite-length ladder strip $\mathcal{S}$ which extends from the plaquette $p_1$ to the plaquette $p_2$, as shown in Fig. 5. These vortex operators are the generalization of 't Hooft vortex operators [10], which create $Z_N$ center flux vortices on a plaquette. We will see that, like the invisible string $\mathcal{L}$ associated with the Wilson charge operators (35), the path $\mathcal{S}$ is also invisible and only the locations of the end points (plaquettes $p_1$ and $p_2$) matter. We now write the SU(2) vortex pair creation operator as

$$\Sigma_{\vec{\omega}}(p_2,p_1) = \exp i\left(\sum_{s=0}^{\mathsf{L}}\hat{\omega}_s\cdot\vec{E}_+(s;\hat{2})\right)\omega. \quad (41)$$

In this equation $E^a_+(s;\hat{2})$ are the left or bottom electric fields on the vertical link $U(s;\hat{2}) \equiv U(x_0+s,y_0;\hat{2}) \in \mathcal{S}$. The axes of rotations $\hat{\omega}_s$ along the ladder strip $\mathcal{S}$ are related to $\hat{\omega}_{s=0} \equiv \hat{\omega}(x_0,y_0) = \hat{\omega}$ by parallel transports

$$\hat{\omega}^a_{h+1} = R^a_b(\mathcal{W}^\dagger(h)U(h;\hat{1}))\hat{\omega}^b_h, \quad h = 0,1,\ldots,\mathsf{L}-1. \quad (42)$$

We have used the notation $U(h;\hat{1}) \equiv U(x_0+h,y_0;\hat{1})$ and $\mathcal{W}(h) \equiv \mathcal{W}(x_0+h,y_0)$, with the plaquette operators $\mathcal{W}(x,y)$ defined in (8). The vortex operator $\Sigma_{\vec{\omega}}(p_2,p_1)$ creates magnetic vortices on the plaquettes $p_1$ and $p_2$ which are located at the end points of the ladder strip $\mathcal{S}$. All other plaquette magnetic fluxes along $\mathcal{S}$ remain zero [7]. Therefore, we can define magnetic vortex states as

$$|\psi_{\vec{\omega}}(p_2,p_1)\rangle_{(\mathsf{p},\mathsf{q})} \equiv \Sigma_{\vec{\omega}}(p_2,p_1)|\psi_0\rangle_{(\mathsf{p},\mathsf{q})}. \quad (43)$$

These are the states with magnetic vortices on the plaquettes $p_1$ and $p_2$ as they satisfy (see the Appendix)

$$B_p|\psi_{\vec{\omega}}(p_2,p_1)\rangle_{(\mathsf{p},\mathsf{q})}$$
$$= \begin{cases} [1-\cos(\frac{\omega}{2})]|\psi_{\vec{\omega}}(p_2,p_1)\rangle_{(\mathsf{p},\mathsf{q})} & \text{if } p=p_1 \text{ or } p_2 \\ |\psi_{\vec{\omega}}(p_2,p_1)\rangle_{(\mathsf{p},\mathsf{q})} & \text{otherwise.} \end{cases}$$
$$(44)$$

The vortex operator $\Sigma_{\vec{\omega}}(p_2,p_1)$ is invariant under (6),

$$A_n|\psi_{\vec{\omega}}(p_2,p_1)\rangle_{(\mathsf{p},\mathsf{q})} = 0 \,\forall\, n, p_1, p_2. \quad (45)$$

Therefore, they are the eigenstates of the toric code Hamiltonian (9),

$$H|\psi_{\vec{\omega}}(p_2,p_1)\rangle_{(\mathsf{p},\mathsf{q})} = 4B\left[1-\cos\left(\frac{\omega}{2}\right)\right]|\psi_{\vec{\omega}}(p_2,p_1)\rangle_{(\mathsf{p},\mathsf{q})}.$$

We note that, unlike the electric charge sector, the magnetic vortex spectrum is continuous. However, a gap can be created in the magnetic sector also by adding an electric-magnetic interaction term in (9) which involves both $A_n$ and $B_p$. For example, using a one-to-one correspondence between lattice sites $n$ and plaquettes $p$ on $\mathcal{T}_2$, we add the following term in the Hamiltonian:

$$\Delta H = C\sum_n(\alpha A_n - B_p)^2. \quad (46)$$

Here $C$ and $\alpha$ are positive constants. The above interaction term leaves the $N^2$ ground states and the exact solvability of the model intact. However, in the large-$C$ limit, the energetically allowed low-energy excitations become discrete and their energies can be manipulated by tuning the parameter $\alpha$.

## IV. NON-ABELIAN ANYONS

After having discussed the $N^2$ degenerate ground states and their electric and magnetic excitations, we now briefly analyze their non-Abelian anyonic nature. As we will see, the non-Abelian nature originates from the basic non-Abelian canonical commutation relations (1). This is also the case with the Kitaev $Z_2$ toric code model. We consider taking an electrically charged particle in the fundamental representation around a loop $\mathcal{C}$ in the presence of magnetic vortices at $p_1$ and $p_2$. This is shown in Fig. 6. Using the canonical commutation relations between the electric fields and the potentials (1), we get

$$\Sigma_{\vec{\omega}}(p_2,p_1)\mathcal{W}_{\alpha\beta}(\mathcal{C})\Sigma^{-1}_{\vec{\omega}}(p_2,p_1)$$
$$= \begin{cases} D^{j=1/2}_{\alpha\gamma}(\hat{\omega}',\omega)\mathcal{W}_{\gamma\beta}(\mathcal{C}) & \text{if } \mathcal{C} \text{ encloses } p_1 \\ D^{j=1/2}_{\alpha\gamma}(\hat{\omega}'',\omega)\mathcal{W}_{\gamma\beta}(\mathcal{C}) & \text{if } \mathcal{C} \text{ encloses } p_2 \\ \mathcal{W}_{\alpha\beta}(\mathcal{C}) & \text{otherwise.} \end{cases} \quad (47)$$

The algebra (47) is the generalized Wilson–'t Hooft order-disorder algebra [7,10] (see the Appendix for details). Thus an electric charge encircling a vortex undergoes SU(2) rotation by the Wigner matrix.[6] This can be seen as follows. We consider the following initial state $|I\rangle$ and the final state $|F\rangle$:

$$|I\rangle^{(\omega)}_{\alpha\beta} \equiv \Sigma_{\vec{\omega}}\Gamma^{(j=1/2)}_{\alpha\beta}(n,n')|\psi_0\rangle_{\mathsf{p},\mathsf{q}},$$
$$|F\rangle^{(\omega)}_{\alpha\beta} \equiv \mathcal{W}_{\alpha\bar{\alpha}}(\mathcal{C})\Sigma_{\vec{\omega}}\Gamma^{(j=1/2)}_{\bar{\alpha}\beta}(n,n')|\psi_0\rangle_{\mathsf{p},\mathsf{q}}. \quad (48)$$

Both the initial and final states are in the (p, q) sector. Using (47) we get

$$|F\rangle^{(\omega)}_{\alpha\beta} = \begin{cases} D^{j=1/2}_{\alpha\bar{\alpha}}(\hat{\omega}',\omega)|I\rangle^{(\omega)}_{\bar{\alpha}\beta} & \text{if } \mathcal{C} \text{ encloses a vortex} \\ |I\rangle^{(\omega)}_{\alpha\beta} & \text{otherwise.} \end{cases} \quad (49)$$

---

[6]In (47) the unit vectors $\hat{\omega}'$ and $\hat{\omega}''$ are related to the unit vector $\hat{\omega}$ through parallel transports which depend on the shapes and sizes of the string $\mathcal{S}$ as well as $\mathcal{C}$. Therefore, axis orientations are unphysical. However, the magnitude $\omega$ is invariant under (6) and is also robust against any deformations or choices of the closed curve $\mathcal{C}$ or $\mathcal{S}$ in Fig. 6.

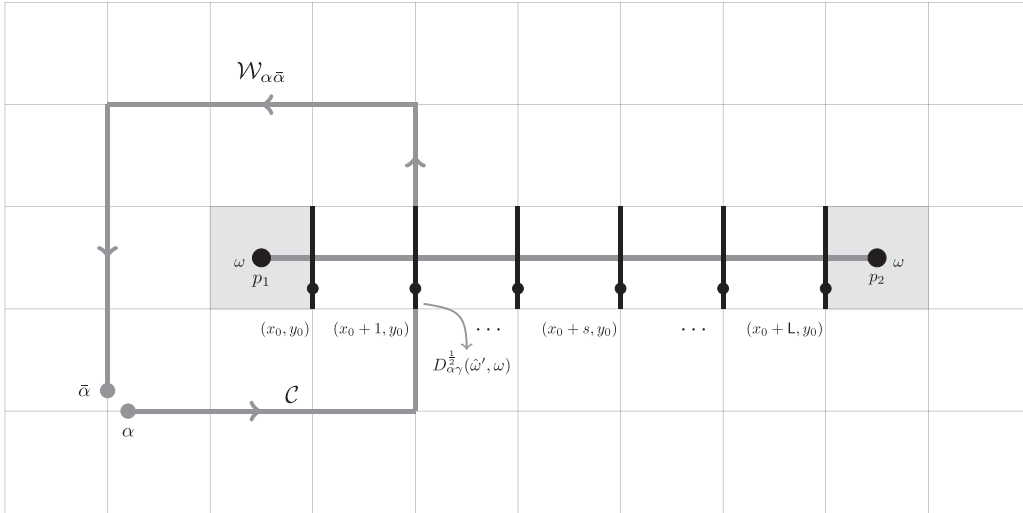

FIG. 6. Taking a charge around closed loop $\mathcal{C}$ produces a non-Abelian rotation by the Wigner rotation matrix if a vortex is enclosed as shown in (49).

The above result is illustrated in Fig. 6. We note that the topological charges (p, q) are not detected by the Wilson loop $\mathcal{W}(\mathcal{C})$ in (49). This is an expected result as no local measurement can distinguish different topological sectors. We can also consider the electric charges in the higher-$j$ representations leading to the Wigner rotation matrices $D^j_{mm'}(\hat{\omega}', \omega)$ as the matrix on the right-hand side of (47). We note that the composites of $\Gamma$ and $\Sigma$ have both electric and magnetic charges and behave like dyons. Their mutual interchange or braiding will lead to non-Abelian anyonic statistics.

## V. CONCLUSION

In this work we have constructed and analyzed the SU(N) toric code model. The $N^2$ topologically stable ground states and all possible electric and magnetic excitations were obtained. The electric charge and magnetic vortex creation-annihilation operators were constructed and their relative non-Abelian anyonic nature was discussed. We expect that these non-Abelian models will also have potential applications in topological quantum computing in the future. In the recent past, there was immense activity to simulate non-Abelian lattice gauge theory Hamiltonians using cold atomic gases and optical lattices [11]. We hope that such attempts will also lead to physical realization of the SU(N) toric code Hamiltonian. The braiding operations on the SU(N) non-Abelian anyons in the spin network Hilbert space should be of interest from the point of view of topological quantum computing.

## ACKNOWLEDGMENT

We would like to thank D. Sen for useful discussions.

## APPENDIX: INVISIBILITY OF DIRAC STRING

In this Appendix we show that only the end points of the Dirac strings are physical where they create the non-Abelian magnetic vortices [7]. We also show that in between the two end points the Dirac strings can be moved around by the non-Abelian gauge transformations. These results also establish the algebra (47). We start with the disorder operator in Fig. 5 with a straight horizontal Dirac string of length $L + 1$. The disorder operator is

$$\Sigma_{\vec{\omega}}(p_2, p_1) = \exp\left(i\omega \sum_{s=0}^{L} \hat{\omega}_s^a \cdot \vec{E}_+^a(s; \hat{2})\right), \quad \text{(A1)}$$

where $\omega = \omega(x_0, y_0) = \omega_{s=0}$ and $E_+^a(s; \hat{2})$, $s = 0, 1, \ldots, L$, are the left or bottom electric fields on the vertical links $U(s; \hat{2}) \equiv U(x_0 + s, y_0; \hat{2})$ which generate SU(2) rotations. The axes of rotations $\hat{\omega}_s$ are related to $\hat{\omega}_{s=0} \equiv \omega(x_0, y_0) \equiv \hat{\omega}$ by parallel transports

$$\hat{\omega}_{h+1}^a = R^{ab}(\mathcal{W}^\dagger(h))U(h; \hat{1})\hat{\omega}_h^b, \quad h = 0, 1, \ldots, L - 1. \quad \text{(A2)}$$

Here $\mathcal{W}(s) \equiv \mathcal{W}(x_0 + s, y_0)$ with the convention defined in (8) and $U(s, \hat{1}) \equiv U(x_0 + s, y_0; \hat{1})$. The parallel transports in the recurrence relations (A2) are the underlying reason for the invisibility of the Dirac strings. This can be seen as follows. The $L + 1$ operators $E^a$ in (A1) rotate the corresponding $L + 1$ vertical links $U(s; \hat{2})$, $0 \leqslant s \leqslant L$, by Wigner $D$ matrices as

$$\Sigma_{\vec{\omega}}(p_2, p_1) \, U_{\alpha\beta}(s; \hat{2}) \Sigma_{\vec{\omega}}^{-1}(p_2, p_1)$$
$$= [D^{j=1/2}(\hat{\omega}_s, \omega)U(s; \hat{2})]_{\alpha\beta}. \quad \text{(A3)}$$

In the above calculation we have used the canonical commutation relations (1). The set of rotated links perpendicular to the Dirac string $\mathcal{S}$ in Fig. 5 do not affect the magnetic fields of any plaquette in the middle of the string but create magnetic vortices at the two end plaquettes $p_1$ and $p_2$. Note that any plaquette in the middle gets its two vertical links rotated around two different axes of rotations. These two axes are related by parallel transports in (A2), so the net effect of rotation is zero and therefore there are no vortex excitations. On the other hand, the two end plaquettes $p_1$ and $p_2$ of the Dirac string $\mathcal{S}$ have only one rotated link each. This results in the creation of two magnetic vortices at $p_1$ and $p_2$. We now consider each of the above three cases separately.

*Case I.* We first consider the plaquette $\mathcal{W}(p_1)$. For the sake of calculations, we write $\mathcal{W} = U_B U_R U_T U_L^\dagger$, where $U_B, U_R, U_T,$ and $U_R$ denote the bottom, right, top, and left links of the plaquette $p_1$ [Eq. (8)]. The disorder operators will rotate only the right link $U_R$ of the plaquette from the left (bottom) end around axis $\hat{\omega}_0$:

$$\Sigma_{\vec{\omega}}(p_2, p_1)\mathcal{W}_{\alpha\beta}(p_1)\Sigma_{\vec{\omega}}^{-1}(p_2, p_1)$$
$$= [U_B D^{j=1/2}(\hat{\omega}_0, \omega) U_R U_T^\dagger U_L^\dagger]_{\alpha\beta}$$
$$= [U_B D^{j=1/2}(\hat{\omega}_0, \omega) U_B^\dagger U_B U_R U_T^\dagger U_L^\dagger]_{\alpha\beta}$$
$$= [U_B D^{j=1/2}(\hat{\omega}_0, \omega) U_B^\dagger \mathcal{W}(p_1)]_{\alpha\beta}.$$

Now we use the following property for the Wigner $D$ function: $U_B D^{1/2}(\hat{\omega}_0^a, \omega) U_B^\dagger = D^{1/2}(R_b^a(U_B^\dagger)\hat{\omega}_0^b, \omega) \equiv D^{1/2}(\hat{\omega}'_0, \omega)$ $[\hat{\omega}'^a_0 \equiv R_b^a(U_B^\dagger)\hat{\omega}_0^b]$. This yields

$$\Sigma_{\vec{\omega}}(p_2, p_1)\mathcal{W}_{\alpha\beta}(p_1)\Sigma_{\vec{\omega}}^{-1}(p_2, p_1)$$
$$= [D^{j=1/2}(\hat{\omega}'_0, \omega)\mathcal{W}(p_1)]_{\alpha\beta}. \quad (A4)$$

*Case II.* Consider a plaquette in the middle of the Dirac string and write $\mathcal{W}(s) = \mathcal{W}(x_0 + s, y_0) = U_B U_R U_T^\dagger U_L^\dagger$, $0 < s < \mathsf{L}$. Now the disorder operator will rotate the two links $U_L^\dagger$ and $U_R$ from the right and the left around axes $\hat{\omega}_s$ and $\hat{\omega}_{s+1}$, respectively, as

$$\Sigma_{\vec{\omega}}(p_2, p_1)\mathcal{W}_{\alpha\beta}(s)\Sigma_{\vec{\omega}}^{-1}(p_2, p_1)$$
$$= [U_B D^{j=1/2}(\hat{\omega}_{s+1}, \omega) U_R U_T^\dagger U_L^\dagger D^{\dagger j=1/2}(\hat{\omega}_s, \omega)]_{\alpha\beta}$$
$$= [U_B D^{j=1/2}(\hat{\omega}_{s+1}, \omega) U_B^\dagger U_B U_R U_T^\dagger U_L^\dagger D^{\dagger j=1/2}(\hat{\omega}_s, \omega)]_{\alpha\beta}$$
$$= [U_B D^{j=1/2}(\hat{\omega}_{s+1}, \omega) U_B^\dagger \mathcal{W} D^{\dagger j=1/2}(\hat{\omega}_s, \omega)]_{\alpha\beta}.$$

Form Eq. (A2) we have $\hat{\omega}_{s+1}^a = R_b^a(\mathcal{W}^\dagger(s)U_B)\hat{\omega}_s$ and

$$\Sigma_{\vec{\omega}}(p_2, p_1)\mathcal{W}_{\alpha\beta}(p)\Sigma_{\vec{\omega}}^{-1}(p_2, p_1)$$
$$= [U_B U_B^\dagger \mathcal{W} D^{j=1/2}(\hat{\omega}_s, \omega) \mathcal{W}^\dagger U_B U_B^\dagger \mathcal{W} D^{\dagger j=1/2}(\hat{\omega}_s, \omega)]_{\alpha\beta}$$
$$= \mathcal{W}_{\alpha\beta}(p).$$

*Case III.* We now consider plaquette $\mathcal{W}(p_2)$ and write $\mathcal{W}(p_2) = U_B U_R U_T^\dagger U_L^\dagger$. The disorder operators will now rotate only the left link $U_L^\dagger$ of $p_2$ from the right around axis $\hat{\omega}_L$:

$$\Sigma_{\vec{\omega}}(p_2, p_1)\mathcal{W}(p_2)\Sigma_{\vec{\omega}}^{-1}(p_2, p_0)$$
$$= [U_B U_R U_T^\dagger U_L^\dagger D^{\dagger j=1/2}(\hat{\omega}_L, \omega)]_{\alpha\beta}$$
$$= [\mathcal{W}(p_2) D^{\dagger j=1/2}(\hat{\omega}_L, \omega)]_{\alpha\beta}$$
$$= [\mathcal{W}(p_2) D^{\dagger j=1/2}(\hat{\omega}_L, \omega) \mathcal{W}^\dagger(p_2) \mathcal{W}(p_2)]_{\alpha\beta}$$
$$= [D^{\dagger j=1/2}(\hat{\omega}'_L, \omega) \mathcal{W}(p_2)]_{\alpha\beta}.$$

In the above we have used $\hat{\omega}'^a_L = R_b^a(\mathcal{W}^\dagger(p_2))\hat{\omega}_L^b$. We can now easily establish the algebra (47) by considering an arbitrary Wilson loop $\mathcal{W}(\mathcal{C})$. If $\mathcal{W}(\mathcal{C})$ encircles one of the two magnetic vortices created by $\Sigma_{\vec{\omega}}(p_2, p_1)$, it will cut the Dirac string $\mathcal{S}$ along a vertical link $U(s, \hat{2}) \in \mathcal{S}$. For such a Wilson loop we can always write $\mathcal{W}(\mathcal{C}) \equiv V_1 U(s; \hat{2}) V_2$ and now

$$\Sigma_{\vec{\omega}}(p_2, p_1)\mathcal{W}(\mathcal{C})\Sigma_{\vec{\omega}}^{-1}(p_2, p_1)$$
$$= V_1 \Sigma_{\vec{\omega}}(p_2, p_1) U(s, \hat{2}) \Sigma_{\vec{\omega}}^{-1}(p_2, p_1) V_2$$
$$= V_1 D^{j=1/2}(\hat{\omega}_s, \omega) U(s, \hat{2}) V_2$$
$$= V_1 D^{j=1/2}(\hat{\omega}_s, \omega) V_1^\dagger V_1 U(s, \hat{2}) V_2$$
$$= D^{j=1/2}(\hat{\omega}', \omega) \mathcal{W}(\mathcal{C}),$$

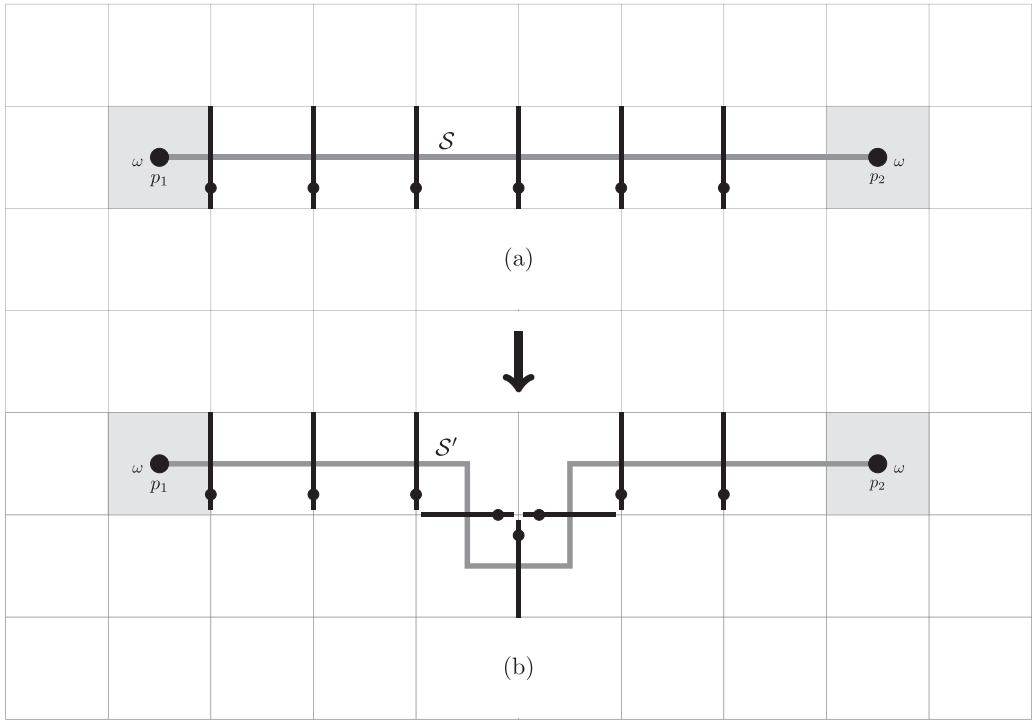

FIG. 7. Deformation of the Dirac string $\mathcal{S}$ by gauge transformations.

where the axis of rotation $\hat{\omega}'^a = R^a_b(V_1)\hat{\omega}^b_s$. Here we note that the axis of rotation also depends upon $V_1$; therefore, is will be different for the two magnetic vortices $p_1$ and $p_2$. If $\mathcal{W}(\mathcal{C})$ does not enclose any magnetic vortex, it will never cut the Dirac string and therefore we get

$$\Sigma_{\vec{\omega}}(p_2, p_1)\mathcal{W}(\mathcal{C})\Sigma^{-1}_{\vec{\omega}}(p_2, p_1) = \mathcal{W}(\mathcal{C}).$$

**Dirac strings and gauge transformations**

We now show that the shape of the Dirac strings between $p_1$ and $p_2$ can be arbitrarily moved around by gauge transformations. We again consider the horizontal straight Dirac string $\mathcal{S}$ in Fig. 7(a). We consider the operator $\mathcal{I}_s = \exp[-i\omega\hat{\omega}^a_s \mathcal{G}^a(s)]$, which will generate gauge transformations at site $(x_0 + s, y_0)$. On any physical state $\mathcal{G}^a|\psi\rangle_{\text{phys}} = 0$, $\mathcal{I}_s$ is an identity operator. Therefore, the rotation on any link $U(s; \hat{2})$ produced by $\Sigma_{\vec{\omega}}(p_2, p_1)$ can be replaced by the rotations produced by $\mathcal{I}_s\Sigma_{\vec{\omega}}(p_2, p_1)$ on the other three links emanating from the same site $s$ resulting in a deformed Dirac string $\mathcal{S}'$. This is shown in Fig. 7(b). Any plaquette other than $p_1$ and $p_2$ will still share two rotated links and therefore will remain unaffected by the new Dirac string.